# Effect of further-neighbor interactions on the magnetization behaviors of the Ising model on a triangular lattice

J. Chen[1], W. Z. Zhuo[1], M. H. Qin[1, a)], S. Dong[2], M. Zeng[1], X. B. Lu[1], X. S. Gao[1], and J. -M. Liu[3, b)]

[1]*Institute for Advanced Materials and Guangdong Provincial Key Laboratory of Quantum Engineering and Quantum Materials, South China Normal University, Guangzhou 510006, China*

[2]*Department of Physics, Southeast University, Nanjing 211189, China*

[3]*Laboratory of Solid State Microstructures, Nanjing University, Nanjing 210093, China*

**[Abstract]** In this work, we study the magnetization behaviors of the classical Ising model on the triangular lattice using Monte Carlo simulations, and pay particular attention to the effect of further-neighbor interactions. Several fascinating spin states are identified to be stabilized in certain magnetic field regions, respectively, resulting in the magnetization plateaus at 2/3, 5/7, 7/9 and 5/6 of the saturation magnetization $M_S$, in addition to the well known plateaus at 0, 1/3 and 1/2 of $M_S$. The stabilization of these interesting orders can be understood as the consequence of the competition between Zeeman energy and exchange energy.

---

a) Electronic mail: qinmh@scnu.edu.cn
b) Electronic mail: liujm@nju.edu.cn

## I. Introduction

During the past decades, the frustrated Ising model on the two-dimensional triangular lattice has attracted widespread interest from both theoretical and experimental approaches, because it can be applied to the description of real materials such as the triangular spin-chain system $Ca_3Co_2O_6$ and Ising magnet $FeI_2$, which usually exhibit fascinating multi-step magnetization behaviors.[1-6] For instance, the anisotropic triangular Ising model has been successfully used to explain the fractional magnetization plateaus experimentally reported in $FeI_2$.[7,8]

As a matter of fact, the triangular Ising model with the antiferromagnetic (AFM) interaction of only nearest neighbor $J_1$ was exactly solved as early as 1950.[9] The spin configurations UUD (spin-up, spin-up, and spin-down) and DDU appear with the same probability in a triangular sublattice, and no spontaneous long-range spin order can be developed at any finite temperature ($T$).[10] The infinite degeneracy can be broken by the introduction of an exchange anisotropy or AFM interaction of second neighbors $J_2$, leading to the stabilization of the collinear AFM state (the spin structure is shown in Fig. 1(a)).[11,12] Furthermore, the UUD state (Fig. 1(b)) is developed when a magnetic field ($h$) is applied, giving rise to the magnetization ($M$) plateau at 1/3 of the saturation magnetization, $M = M_S/3$. The $M_S/3$ plateau persists up to the saturation field, resulting in the two-step magnetization behaviors[13] which have been experimentally reported in several classical and quantum triangular spin systems.[14-17] In our earlier work, the ground-state two-step magnetization behavior has been further confirmed by the Wang-Landau simulation of the model even with the random-exchange interaction.[18]

Interestingly, $J_2$ interaction is proved to be very efficient in modulating the magnetization behaviors in such a frustrated spin system because of the effective reduction of the nearest neighboring interaction due to the frustration.[19] In detail, the inclusion of the $J_2$ term produces the $M_S/2$ magnetization plateau, in addition to the $M_S/3$ plateau. The $M_S/2$ plateau is caused by the ferrimagnetic (FI) state with spin arrangement consisting of alternating AFM and ferromagnetic (FM) stripes, as shown in Fig. 1(c).[20] It is noted that every spin-down in the UUD state is antiparallel to its nearest neighbors, and these local $J_1$ interactions are satisfied, while all the $J_2$ interactions in the whole system are dissatisfied. On the other hand, all the AFM $J_1$ and $J_2$ interactions between every down-spin site and its neighbors are satisfied in the FI state. Thus, more local AFM $J_2$ interactions are satisfied in the FI state, leading to the replacement of the $M_S/3$ plateau by the $M_S/2$ one with the increasing $J_2$. In addition, the energy loss from the $J_2$ interaction due to the phase transition from the FI state to the FM state linearly increases with $J_2$, leading to the further increase of the width of the 1/2 plateau.

In some extent, the same mechanism for the stabilization of the FI state by AFM $J_2$ may also work for the system with further-neighbor interactions, and more interesting plateaus states may be available. The importance of the study on this subject can be explained in the following two aspects. On one hand, several interesting plateaus have been reported in some of the triangular antiferromagnets.[21,22] For example, the magnetization plateaus at 1/3, 1/2, and 2/3 of $M_S$ are experimentally reported in $Ba_3CoNb_2O_9$. The first plateau state is believed to originate from the easy-axis anisotropy, while the other two plateaus states are still not very clear.[21] It is noted that additional interactions may be available in real materials and may play an important role in modulating the magnetization behaviors. In fact, a narrow 2/3 magnetization plateau has been predicted when the dipole-dipole interaction is considered in the model for Shastry-Sutherland magnets $TmB_4$, and one may question that if the same mechanism still holds true for triangular antiferromagnets.[23] On the other hand, the study of these nontrivial magnetic orders also contributes to the development of statistical mechanics. For example, the disordering of the collinear phase in the triangular Ising model with distant neighboring interactions has been proved to occur via a first-order phase transition.[24,25] However, the role of further-neighbor interactions in modulating the magnetization behaviors in triangular Ising antiferromagnets is far from being completely understood, which deserves to be checked in detail.

In this work, as the first step, we study the classical Ising model with further-neighbor interactions on the triangular lattice, as shown in Fig. 1(d). Several interesting spin orders are identified to be developed in certain $h$ regions, respectively, leading to the emergence of the plateaus at 2/3, 5/7, 7/9 and 5/6 of $M_S$, in addition to the plateaus at 0, 1/3 and 1/2 of $M_S$. Furthermore, the magnetic structures of the 1/2 and 2/3 plateaus states uncovered in this work are consistent with the earlier predictions for $Cs_2CuBr_4$, demonstrating the generality of these states, at least, partially.[19]

The rest of this paper is organized as follows. In Sec. II, the model and the simulation method will be presented and described. Section III is attributed to the simulation results and discussion. The conclusion is presented in Sec. V.

## II. Model and method

In the presence of $h$ and further-neighbor interactions, the model Hamiltonian can be described as follows:

$$H = J_1 \sum_{\langle ij \rangle_1} S_i \cdot S_j + J_2 \sum_{\langle ij \rangle_2} S_i \cdot S_j + J_3 \sum_{\langle ij \rangle_3} S_i \cdot S_j + J_4 \sum_{\langle ij \rangle_4} S_i \cdot S_j + J_5 \sum_{\langle ij \rangle_5} S_i \cdot S_j - h \sum_i S_i, \quad (1)$$

where $J_1 = 1$ is the unit of energy, $S_i$ represents the Ising spin with unit length on site $i$, $\langle ij \rangle_n$ ($n$ = 1, 2, 3, 4, 5) denotes the summation over all pairs on the bonds with $J_n$ coupling as shown in Fig. 1(d). Besides, all of distant neighbors are antiferromagnetic coupled and with $J_n/J_{n+1} > 1.4$ ($n$ = 1, 2, 3, 4), to be consistent with the dipole-dipole interaction, in some extent.

Furthermore, such a frustrated spin system may be easily trapped into metastable states at low $T$ and is hard to relax to the equilibrium state. To overcome this difficulty, one may appeal to the parallel tempering exchange Monte Carlo (MC) method which efficiently prevents the system from trapping in metastable free-energy minima caused by the frustration.[26,27] Our simulation is performed on an $N = L \times L$ ($L$ is reasonably chosen to be consistent with the size of unit cell) triangular lattice with period boundary conditions. We take an exchange sampling after every 10 standard MC steps. The simulation is started from the FM state at a high $h$, and the $M(h)$ curves are calculated upon $h$ decreasing. Typically, the initial $2 \times 10^5$ MC steps are discarded for equilibrium consideration and another $2 \times 10^5$ MC steps are retained for statistic averaging of the simulation.

## III. Simulation results and discussion

First, we study the effect of $J_3$ on the magnetization behaviors of the model. Fig. 2(a)

shows the simulated magnetization curves for various $J_3$ with $L = 42$ at $(J_2, J_4, J_5) = (0.1, 0, 0)$ and $T = 0.01$. For $J_3 = 0$, the magnetization plateaus at $M/M_S = 0$, 1/3, and 1/2 are observed.[28] The first collinear-state plateau is gradually melted when $J_3$ increases from zero, demonstrating that the collinear state is not favored by $J_3$.[29] In addition, the 1/3 plateau is broadened at the expense of the 1/2 one with the increase of $J_3$. More interestingly, an additional plateau at $M/M_S = 5/7$ resulted from a particular state with the spin configuration shown in Fig. 2(b) is developed. In the 5/7 state, for every spin-down, all the AFM $J_1$, $J_2$, and $J_3$ interactions are satisfied. Furthermore, all the spins-down form a hexagonal close packed structure to save the exchange energy. The central spin-down may flip as the magnetic energy increases to be comparable with the interaction energy, and the saturation $h$ is estimated to be $6(J_1 + J_2 + J_3)$, well consistent with our simulation.

To understand the simulated results, we calculate the $h$-dependence of the individual energy components of the $J_n$ coupling $E_n$ ($n = 1, 2, 3$) and the Zeeman energy $E_{zee}$ for $J_3 = 0.03$ (Fig. 2(c)). It is noted that the $J_3$ interactions between the down-spin site and its third neighbors in the UUD state are satisfied, while those in the collinear/FI state are not satisfied, leading to the broadening of the 1/3 plateau with the increasing $J_3$ accompanied by the destabilizations of the collinear and 1/2 plateaus. In the 5/7 state, totally $6N/7$ $J_3$ bonds are satisfied, and the enhancement of this state with $J_3$ can be understood from the following two aspects. On one hand, within a certain $h$ range, the energy loss from $E_1$ and $E_2$ due to the phase transition from the FI state to the 5/7 state can be covered by the energy gain from $E_3$ and $E_{zee}$, resulting in the stabilization of the 5/7 state. Furthermore, the energy gain from $E_3$ linearly increases with $J_3$, leading to the gradually replacement of the 1/2 plateau by the 5/7 one. On the other hand, the energy loss from $E_3$ due to the phase transition from the 5/7 state to the FM state increases when $J_3$ is increased, and a larger $h$ is needed to flip spins-down in the 5/7 state. As a result, the $h$-region with the 5/7 state is further enlarged at the expense of that with the FM state.

One may note that the configuration of the 5/7 state is filled by two kinds of hexagonal clusters (connected by red lines in Fig. 2(b)), in which every site belongs to one hexagon as the central one and to six other hexagons as the lateral one. As a matter of fact, the configuration has been exactly proved to be the unique ground state by the "$m$-potential methods", while a detail ground-state phase diagram is not available.[30] Fig. 3 gives the calculated ground-state phase diagram at $J_2 = J_1/10$. Specifically, at the six multi-phase points (A, B, C, D, E, F) confirmed by the MC simulations, we figure out that the hexagonal clusters are with the minimal energy (equations of the calculation are presented in Appendix). In each triangular region (ABC, BCD, etc), the configuration of the ground state is constructed by the hexagonal clusters which are simultaneously existed in all three points. Thus, the collinear state, UUD state, FI state, and 5/7 state are exactly proved to be the ground states, respectively. The phase diagram is well consistent with the MC simulated one (the transition points are estimated from the magnetization jumps) at low $T$ ($T = 0.01$),[31] although a tiny bifurcation is noticeable due to the unavoidable thermal fluctuations in MC simulations, as shown in Fig. 2(d).

Subsequently, we paid attention to the effect of AFM $J_4$ on the magnetization behaviors. The simulated magnetization curves for various $J_4$ with $L = 36$ at $(J_2, J_3, J_5) = (0.2, 0.1, 0)$ and $T = 0.01$ are given in Fig. 4(a), and the corresponding phase diagram in the $J_4$-$h$ plane for $J_4 > 0.025$ is shown in Fig. 4(b) (the transition $h$ at zero $T$ obtained by comparing the energies of these phases are also given with the dotted lines). The phase diagram shows that the width of the 1/3 plateau is increased with the increase of $J_4$, demonstrating that $J_4$ favors the UUD state. Furthermore, two additional plateaus at $M/M_S = 2/3$ and 7/9 are subsequently stabilized with increasing $h$ for $J_4 > 0.025$. The spin configurations of the 2/3 and 7/9 states are shown in Fig. 4(c) and 4(d), respectively. Actually, the 2/3 magnetization step has been observed in several triangular antiferromagnets, and the spin configurations remain to be checked. Here, the spin structure of the 2/3 state uncovered in this work is the same as that predicted for the spin-1/2 Heisenberg system $Cs_2CuBr_4$,[19] demonstrating the common feature of this state, in some extent. Furthermore, for every spin-down in the 7/9 state, all the AFM $J_1$, $J_2$, $J_3$, and $J_4$ interactions are satisfied, and the saturation field $h = 6(J_1 + J_2 + J_3 + 2J_4)$ increases with the increase of $J_4$. In addition, the slopes of the transition fields to the FI, 2/3, and 7/9 states vs $J_4$ are with a fixed value, leading to the fact that the widths of the plateaus at 1/2, 2/3, and 7/9 are invariant for various $J_4$ ($J_4 > 0.025$).

Finally, the effect of AFM $J_5$ is investigated. Fig. 5(a) gives the magnetization curves for various $J_5$ for $L = 48$ at $(J_2, J_3, J_4) = (0.2, 0.1, 0.07)$ at $T = 0.01$. In the $h$ region just below the saturation field, the 7/9 plateau is gradually replaced by the 5/6 plateau when $J_5$ is increased from 0.01. Similarly, for every spin-down in the 5/6 state (Fig. 5(b)), all the AFM $J_1$, $J_2$, $J_3$, $J_4$, and $J_5$ interactions are satisfied. The widths of the 5/6 and 2/3 plateaus are almost the same, and show little dependence on $J_5$. Furthermore, the width of the 1/2 plateau is broadened at the expense of the 1/3 plateau, indicating that $J_5$ favors the 1/2 state more than the UUD state.

Following our earlier work, the transition fields at zero $T$ between the different phases uncovered in our simulations can be also quantitatively obtained by comparing the energies of these phases.[32] In detail, the energy per site of the collinear state, the UUD state, the 1/2 state, the 2/3 state, the 5/7 state, the 7/9 state, the 5/6 state, and the FM state can be exactly calculated based on their spin configurations of the unit cells, respectively, and are stated as follows:

$$H_{\text{collinear}} = -J_1 - J_2 + 3J_3 - 2J_4 - J_5, \tag{2}$$

$$H_{\text{UUD}} = -J_1 + 3J_2 - J_3 - 2J_4 + 3J_5 - \frac{1}{3}h, \tag{3}$$

$$H_{\text{FI}} = 3J_3 - \frac{1}{2}h, \tag{4}$$

$$H_{2/3} = J_1 + J_2 + 2J_3 + 2J_4 + J_5 - \frac{2}{3}h, \tag{5}$$

$$H_{5/7} = \frac{9}{7}J_1 + \frac{9}{7}J_2 + \frac{9}{7}J_3 + \frac{30}{7}J_4 + \frac{9}{7}J_5 - \frac{5}{7}h, \tag{6}$$

$$H_{7/9} = \frac{5}{3}J_1 + \frac{5}{3}J_2 + \frac{5}{3}J_3 + \frac{10}{3}J_4 + 3J_5 - \frac{7}{9}h, \tag{7}$$

$$H_{5/6} = 2J_1 + 2J_2 + 2J_3 + 4J_4 + 2J_5 - \frac{5}{6}h, \tag{8}$$

$$H_{\text{FM}} = 3J_1 + 3J_2 + 3J_3 + 6J_4 + 3J_5 - h, \tag{9}$$

Fig. 5(c) shows the calculated local energies as a function of $h$ for these states at $(J_2, J_3, J_4, J_5) = (0.2, 0.1, 0.07, 0.02)$. Both the 5/7 and 7/9 states are not stabilized in the whole $h$ region, and $H_{5/7}$ and $H_{7/9}$ are not shown for simplicity. Clearly, five transition fields are recognized and exactly calculated, well consistent with the simulated ones, respectively. The width of the 2/3 plateau is calculated to be $6J_3$, which are irrelevant to $J_4$ and $J_5$, as confirmed in our simulations. In some extent, it is suggested that the 2/3 plateau is with a special configuration and can be stabilized by further neighbor AFM interactions.

Despite many years of fruitful research, triangular antiferromagnets continue to offer us with novel spin states and magnetization behaviors which can be efficiently modulated by further-neighbor interactions, as revealed again in this work. Some of these states are with very close local energies, and are very sensitive to external fluctuations. As a result, even a weak additional interaction may have a prominent influence on the magnetization behaviors of the system. Furthermore, all the neighboring interactions studied in this work are antiferromagnetic, which may be available in real materials with strong dipole-dipole interaction.[33] Thus, this work strongly suggests that the interactions between distant neighbors may contribute to the magnetization plateaus observed in real materials, although not all the spin states predicted here have been experimentally reported.

## IV. Conclusion

In conclusion, we have studied the magnetization behaviors and competing spin orders of classical Ising model with interactions between distant neighbors on a triangular lattice by means of Monte Carlo simulations. In addition to the well known plateaus at 0, 1/3 and 1/2 of $M_S$, those at 2/3, 5/7, 7/9 and 5/6 of $M_S$ are predicted when further-neighbor interactions are considered. These fascinating plateaus states are discussed in details and confirmed by the ground state analysis. It is suggested that even weak distant neighboring interactions may have a significant effect on the magnetization behaviors, and may contribute to the experimentally reported magnetization plateaus.

## Acknowledgements

This work was supported by the Natural Science Foundation of China (51322206, 11274094, 51332007), and the National Key Projects for Basic Research of China (2015CB921202 and 2015CB654602).

## Appendix

In this section, we briefly present the process of the "$m$-potential" calculations, and more details can be found in section II of Ref. 30.[30] Considering $J_1$, $J_2$ and $J_3$ interactions, the Hamiltonian can be states as:

$$H = V_1 \sum_{\langle ij \rangle_1} (a_1\sigma_i + b_1)(a_1\sigma_j + b_1) + V_2 \sum_{\langle ij \rangle_2} (a_2\sigma_i + b_2)(a_2\sigma_j + b_2) + V_3 \sum_{\langle ij \rangle_3} (a_3\sigma_i + b_3)(a_3\sigma_j + b_3) - \mu \sum_i (a_0\sigma_i + b_0) \quad (A1)$$

with $a_n = 1$, $b_n = 0$ ($n = 0, 1, 2, 3$). In equation A1, the following notations are introduced:

$$V_n = \frac{J_n}{a_n^2}, n = 1, 2, 3, \quad (A2)$$

$$S_i = a_n\sigma_i + b_n, \ n = 1, 2, 3 \quad (A3)$$

and

$$\mu = \frac{1}{a_0}(h + 6a_1b_1J_1 + 6a_2b_2J_2 + 6a_3b_3J_3). \quad (A4)$$

For an infinite size, the Hamiltonian can be rewritten as:

$$H = \sum_i \frac{V_1}{2(\beta_1 + \beta_2)}[\beta_1\sigma_{i0}^1(\sigma_{i1}^1 + \sigma_{i2}^1 + \sigma_{i3}^1 + \sigma_{i4}^1 + \sigma_{i5}^1 + \sigma_{i6}^1) + \beta_2(\sigma_{i1}^1\sigma_{i2}^1 + \sigma_{i2}^1\sigma_{i3}^1 + \sigma_{i3}^1\sigma_{i4}^1 + \sigma_{i4}^1\sigma_{i5}^1 + \sigma_{i6}^1\sigma_{i1}^1)] +$$
$$\frac{V_2}{2}(\sigma_{i1}^2\sigma_{i3}^2 + \sigma_{i3}^2\sigma_{i5}^2 + \sigma_{i5}^2\sigma_{i1}^2 + \sigma_{i2}^2\sigma_{i4}^2 + \sigma_{i4}^2\sigma_{i6}^2 + \sigma_{i6}^2\sigma_{i2}^2) +$$
$$V_3(\sigma_{i1}^3\sigma_{i4}^3 + \sigma_{i2}^3\sigma_{i5}^3 + \sigma_{i3}^3\sigma_{i6}^3) -$$
$$\frac{\mu}{\alpha_1 + 6\alpha_2}[\alpha_1\sigma_{i0}^0 + \alpha_2(\sigma_{i1}^0 + \sigma_{i2}^0 + \sigma_{i3}^0 + \sigma_{i4}^0 + \sigma_{i5}^0 + \sigma_{i6}^0)] \quad (A5)$$

with

$$\sigma_{ij}^n = a_n\sigma_{ij} + b_n, n = 1, 2, 3 \quad (A6)$$

In detail, the hexagonal configurations at the multiphase points can be obtained from the following equations, respectively:

(1) Point A:

$$h - 12J_2 + 12J_3 = 0, \text{ with } J_3 = 0, \ \alpha_1 = 0, \ \frac{\beta_1}{\beta_2} = 2, \ \frac{a_1}{b_1} = 3 \quad (A7)$$

(2) Point B:

$$h - 6J_1 + 18J_2 - 24J_3 = 0, \text{ with } J_3 = 0, \ \alpha_1 = 0, \ \frac{\beta_1}{\beta_2} = \frac{3}{7}, \ \frac{a_1}{b_1} = 3 \quad (A8)$$

(3) Point C:

$$h - 6J_1 + 18J_2 - 24J_3 = 0, \text{ with } J_3 = 0.03, \ \alpha_1 = 0, \ \frac{\beta_1}{\beta_2} = \frac{7}{3}, \ \frac{a_1}{b_1} = \frac{5}{4} \tag{A9}$$

(4) Point D:

$$h - 6J_1 - 6J_2 + 8J_3 = 0, \text{ with } J_3 = 0.03, \ \frac{\alpha_1}{\alpha_2} = 2, \ \frac{\beta_1}{\beta_2} = \frac{2}{3}, \ b_1 = 0, \ \frac{a_2}{b_2} = \frac{1}{3} \tag{A10}$$

(5) Point E:

$$h - 6J_1 - 6J_2 + 8J_3 = 0, \text{ with } J_3 = 0, \ \alpha_1 = 0, \ \frac{a_1}{b_1} = 1 \tag{A11}$$

(6) Infinite point F:

$$h - 6J_1 - 6J_2 - 6J_3 = 0, \text{ with } h \to \infty, \ \alpha_1 = 0, \ \frac{a_1}{b_1} = 1 \tag{A12}$$

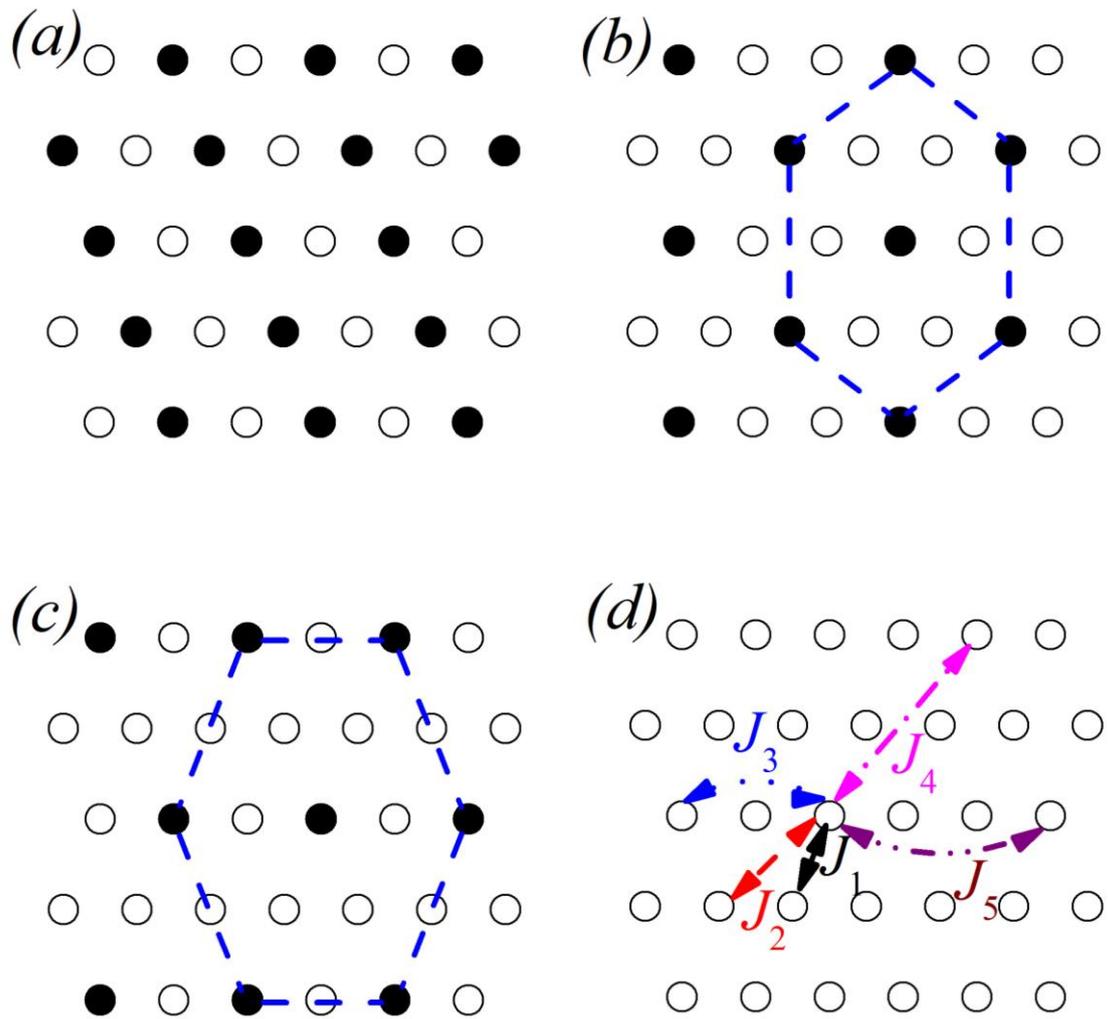

Fig.1. (color online) Spin configurations in (a) the collinear state, (b) the UUD state, and (c) the 1/2 state. Solid and empty circles represent the spins-down and the spins-up, respectively. All the spins-down in the UUD (1/2) state form the hexagonal structure to save the exchange energy. (d) Simulated model on the triangular lattice with the first-, second-, third-, forth- and fifth-nearest neighbor exchange interactions.

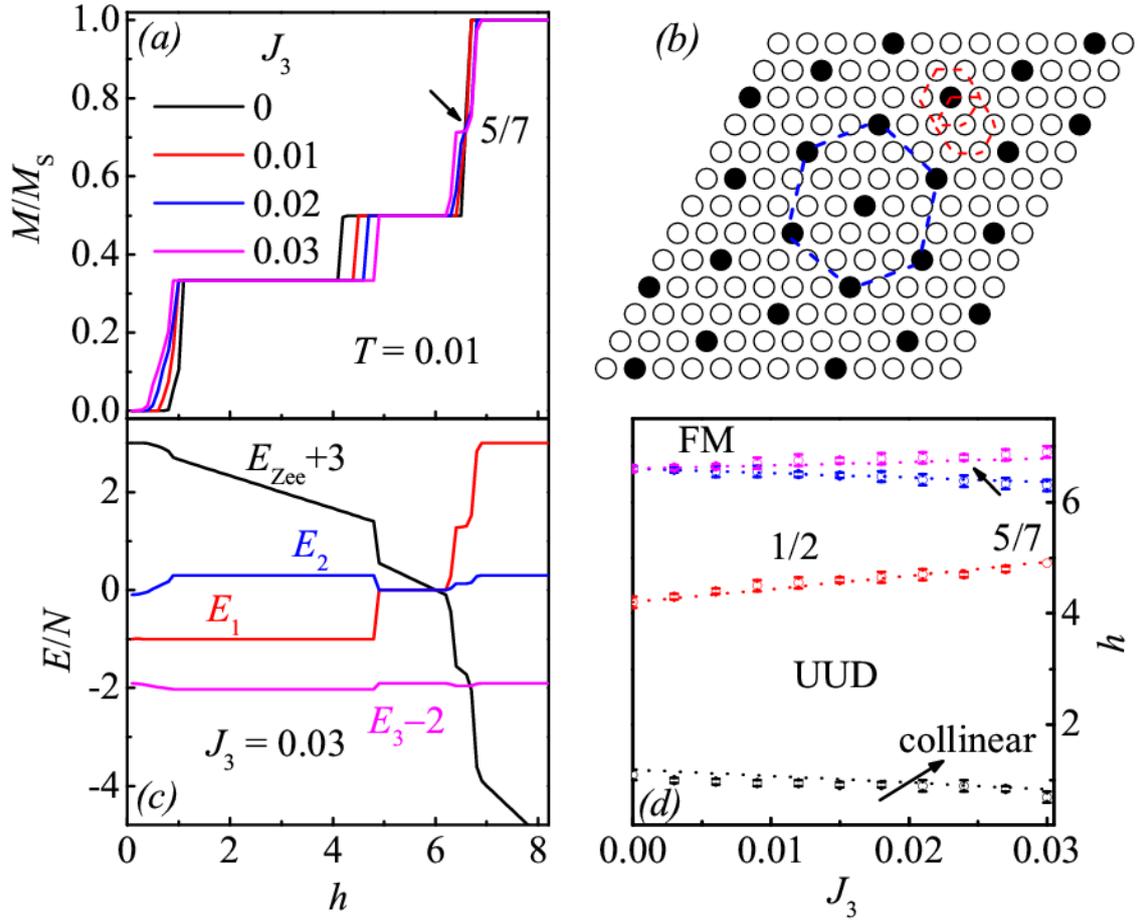

Fig.2. (color online) (a) Magnetization curves for various $J_3$. The parameters are $L = 42$, $T = 0.01$ and $(J_2, J_4, J_5) = (0.1, 0, 0)$. (b) Spin configuration for the state with the plateau at 5/7 of $M_S$. (c) The calculated $E_1$, $E_2$, $E_3-2$, and $E_{zee}+3$ as a function of $h$ at $T = 0.01$ for $J_3 = 0.03$. (d) MC simulated (empty cycles) phase diagram at $T = 0.01$ in the $h$-$J_3$ plane, and the exactly ground-state boundaries are also shown with the dotted lines.

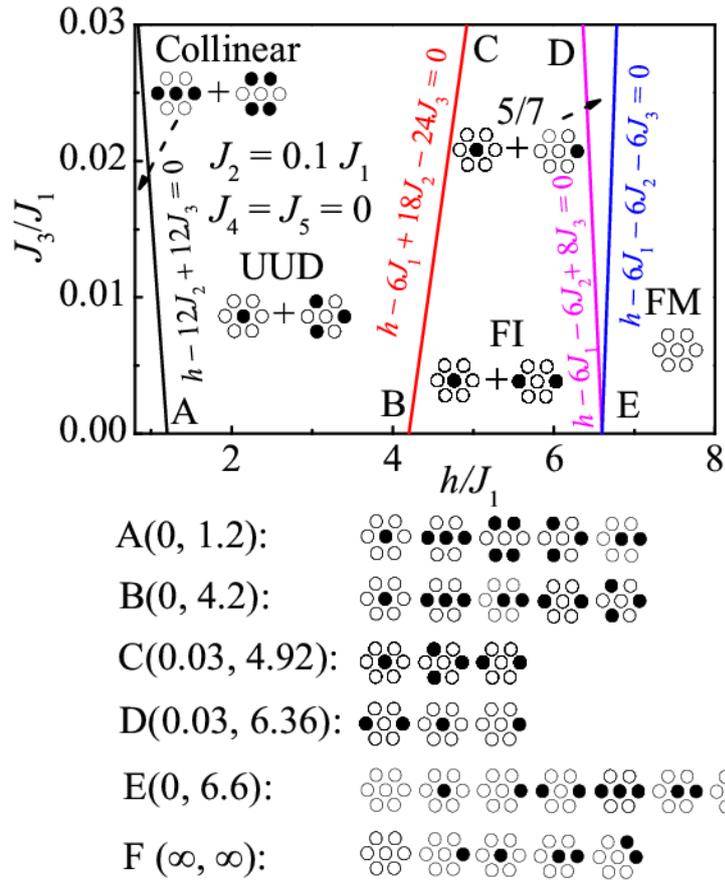

Fig.3. (color online) Ground-state phase diagram in the ($h$, $J_3$) plane at $J_4 = J_5 = 0$ obtained by the "$m$-potential methods". The spin configurations of hexagons are also presented.

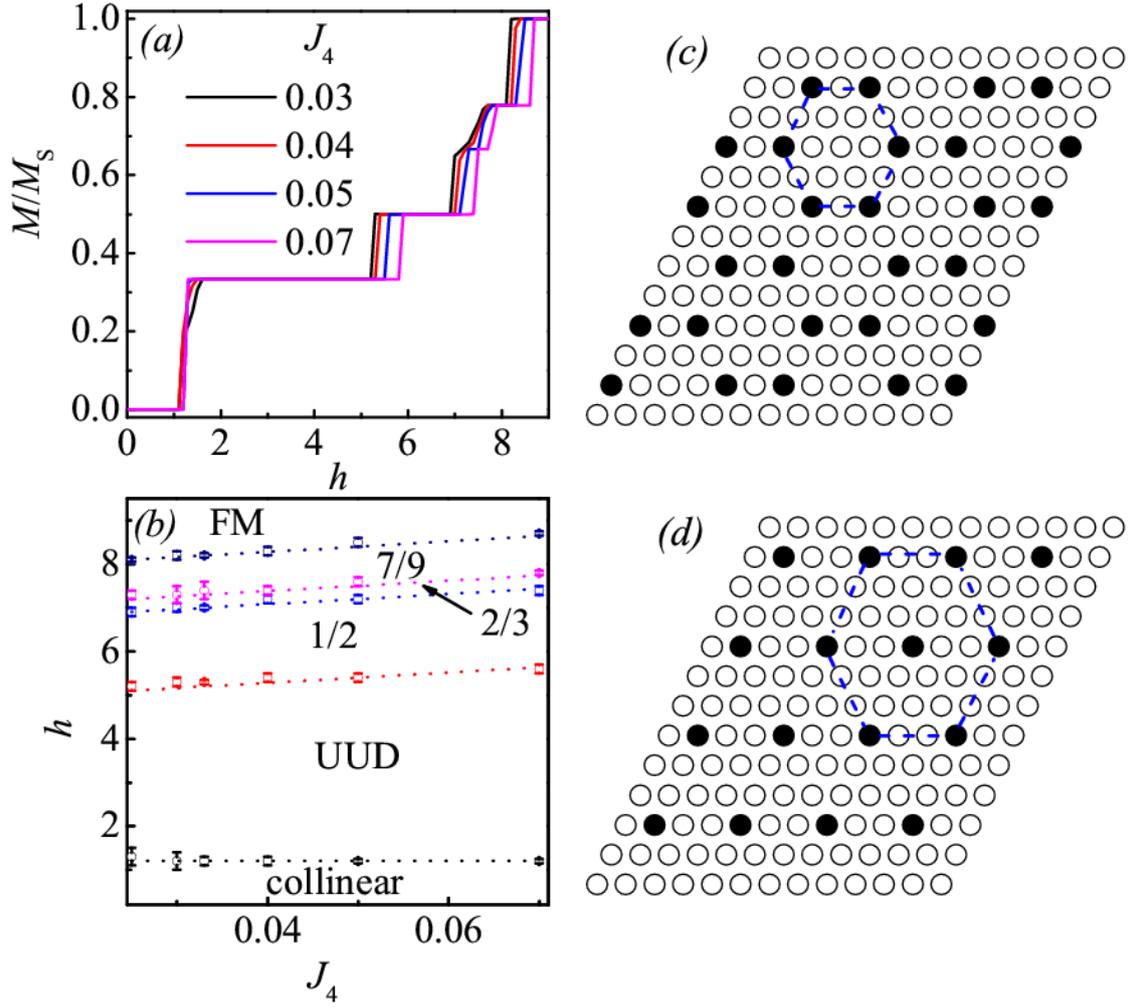

Fig.4. (color online) (a) Magnetization curves for various $J_4$ ($J_4 > 0.025$). The parameters are $L = 36$, $T = 0.01$ and $(J_2, J_3, J_5) = (0.2, 0.1, 0)$. (b) MC simulated (empty cycles) phase diagram at $T = 0.01$ in the $h$-$J_4$ plane ($J_4 > 0.025$), and the exactly ground-state boundaries are also shown with the dotted lines. Spin configurations in the (c) 2/3 plateau state, and (d) 7/9 plateau state.

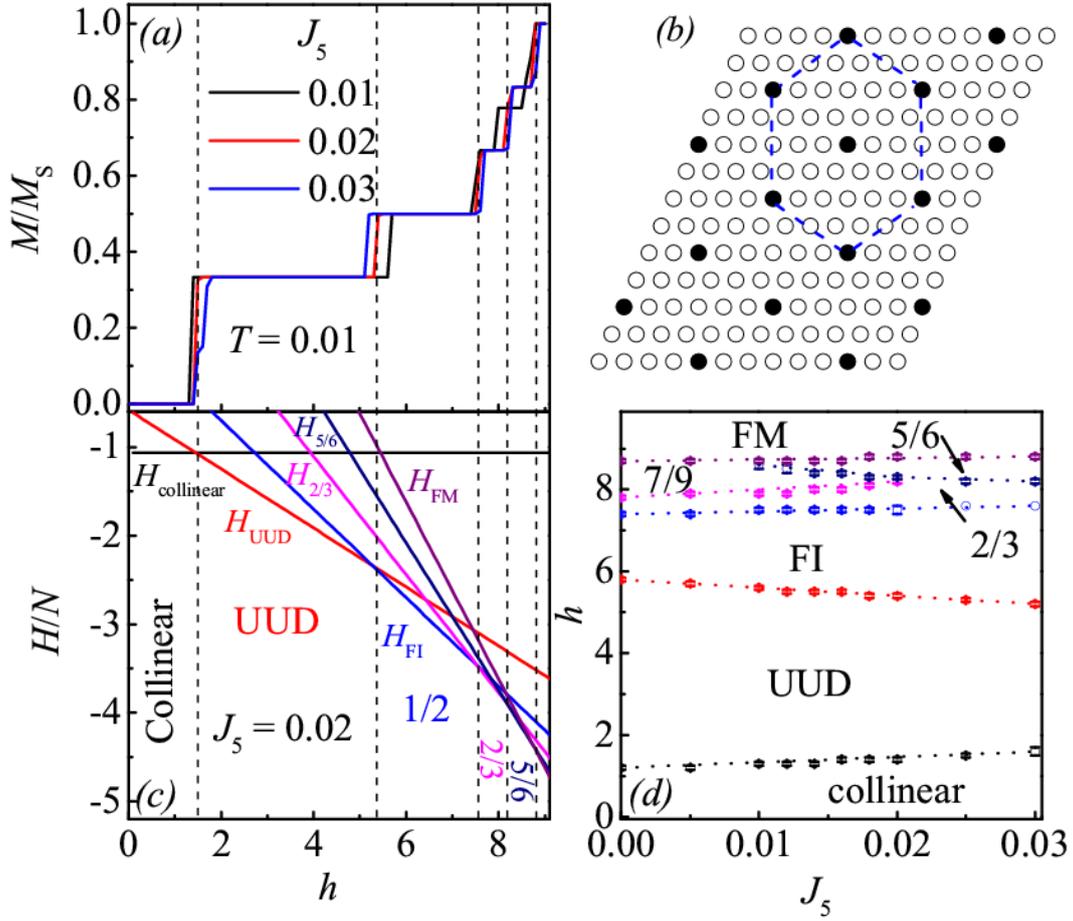

Fig.5. (color online) (a) Magnetization curves for various $J_5$. The parameters are $L = 48$, $T = 0.01$ and $(J_2, J_3, J_4) = (0.2, 0.1, 0.07)$. (b) Spin configurations in the 5/6 plateau state. (c) The local energies as a function of $h$ for $J_5 = 0.02$. (d) MC simulated (empty cycles) phase diagram at $T = 0.01$ in the $h$-$J_5$ plane, and the exactly ground-state boundaries are also shown with the dotted lines.